\documentclass[referee]{aa}
\usepackage{graphicx}

\begin{document}
\title{Thick disks and halos of spiral galaxies M~81, NGC~55 and  NGC~300.
\thanks{Based on observations with the NASA/ESA Hubble Space Telescope,
obtained at the Space Telescope Science Institute, which is operated by the
Association of Universities for Research in Astronomy, Inc., under NASA
contract NAS 5-26555.}}
\titlerunning{Galaxy M~81, NGC~55 and  NGC~300}
\authorrunning{Tikhonov N.A. et al.}
\author{Tikhonov N.A. \inst{1,2}, Galazutdinova O.A. \inst{1,2}, Drozdovsky I.O. \inst{3,4}}
\institute{Special Astrophysical Observatory, Russian Academy
   of Sciences, N.Arkhyz, KChR, 369167, Russia
\and Isaac Newton Institute of Chile, SAO Branch, Russia
\and Spitzer Science Center, Caltech, MC 220-6, Pasadena, CA 91125, USA
\and Astronomical Institute, St.Petersburg University, 198504, Russia}
\date {Received \today }
\abstract
{By using images from the HST/WFPC2/ACS archive, we have analyzed the
spatial distribution of the AGB and RGB stars along the galactocentric radius
of nearby spiral galaxies M~81, NGC~300 and NGC~55.
Examining color-magnitude diagrams and stellar luminosity functions, we gauge
the stellar contents of the surroundings of the three galaxies.
The red giant population (RGB) identified at large galactocentric radii yields
a distance of $3.85\pm0.08$~Mpc for M~81, $2.12\pm0.10$~Mpc for NGC~55,
and $2.00\pm0.13$~Mpc for NGC~300, and a mean stellar metallicity of $-0.65$,
$-1.25$, and $-0.87$ respectively. We find that there are two number density gradients
of RGB stars along the radius, which correspond to the thick disk and halo
components of the galaxies. We confirm the presence of a metallicity gradient
of evolved stars in these galaxies, based on the systematic changes of the
color distribution of red giant stars.
These results imply that the thick disk might be a general feature of
spiral galaxies, and endorse a further investigation of the outer stellar
edges of nearby spirals, which is critical in constraining the origin
and evolution of galaxies.
}
\maketitle

\section{Introduction}

The fossil record of galaxy formation and evolution is imprinted
on the spatial distribution, ages and metallicities of galactic stellar
populations.
The properties of the outer parts of galaxies are very
sensitive to both galaxy star formation history and the nature of the
intergalactic medium.

The various studies of the nearest spiral galaxies, such as the Milky Way, M31
and M33, have revealed their similar stellar structure: bulge, thin and
thick disks and halo (van der Marel, 2001; Zocalli et al., 2002;
Sarajedini \& van Duyne, 2001; Pritchet \& van der Bergh, 1988; Belazzini et al., 2003;
Guillandre et al., 1998; Brown et al., 2003; Zucker et al., 2004).
Aside from the spatial distribution law and kinematics,
stars from each of these subsystems share a common star formation history
 retaining considerable age and chemical information
(Vallenary et al., 2000; Prochaska et al., 2000; Chiba \& Beers, 2000;
Zoccali et al., 2002; Williams, 2002, Sarajedini et al., 2000,
Brewer et al., 2003).
The majority of the investigations concentrated predominantly on the central
parts of these galaxies, biased towards high surface brightness star forming
regions. In recent years,
wide-field observations of the nearest spirals have shown that
they are substantially more extended than previously thought.
The information on the actual origin of these outer regions is scarce,
and has been a matter of debate.
Today, we know that
most of these extended, elusive stellar components are predominantly
evolved, but whether this is a true single burst ancient stellar population,
with little or no intermediate-age component, has not yet been
established. By 'population', we mean here an ensemble of stars that share a
coherent history.

Various tests of formation scenarios have centered on two extreme
viewpoints --- (1) the monolithic collapse model for the galaxy halo
(Eggen et al., 1962) together with internal chemodynamical evolution for
the thick disk (e.g. Burkert et al., 1992), and (2) the accretion model
for both the halo (Searle \& Zinh, 1978) and the thick disk
(Carney et al., 1989; Gilmore, Wyse and Norris 2002).
In view of the merging of the Sagittarius dwarf galaxy with Milky
Way (Ibata, Gilmore \& Irwin 1995), and the predictions of cold dark matter
(CDM) cosmological models for hierarchical galaxy formation from
large number of mergers, the accretion origin of the thick disk and halo is
especially relevant. At the same time, the
significant difference in the chemical abundance pattern of
the Milky Way's
halo stars and its satellite galaxies likely implies a different
evolution history (Brewer \& Carner, 2004; Venn et al., 2004).
An alternative disk thickening model by the evolution of massive
star clusters in a thin disk has been suggested (Kroupa 2002).
By studying the spatial distribution of evolved stars, we may hope to
attain a basic understanding of the general morphology of a galaxy at
the time when the dominant component of its old stellar population formed
and constrain the proposed evolution models.

It is easier to separate the spherical bulge from the exponential
thin disk components based on their distinctive surface brightness profiles
(Kent, 1985; Byun \& Freeman, 1995; Bagget et al., 1998; Prieto et al., 2001)
than the thick disk from halo, due to the extremely low
surface brightness of these galaxy components (typically below the
$\mu_V\sim26$~mag$/\sq\arcsec$).
While some of the recent studies of the outer regions of spiral galaxies
rely on multicolor surface photometry,
successful detection is only possible in the nearest galaxies or
in the edge-on galaxies with a bright and extended thick disk/halo
(e.g. Harris \& Harris, 2001; Dalcanton \& Bernstein, 2002;
Pohlen et al. 2004).
For example, Dalcanton \& Bernstein (2002) analyzed sample of 47
edge-on spiral galaxies and found the  presence
of a thick disk in  90\%  of them, based on the results of multicolor
surface photometry. The other important  result obtained in those studies
is that both thin and thick disks are truncated, with mean scaleheight
and scalelength of the thick disk several times larger than that of
the thin disk (Pohlen et al. 2004).
The question of whether these thick disk structures consist of
evolved stellar populations is rather unclear, due to
ambiguity of the surface brightness colors.
For example, AGB and  RGB stars with different ages
can have almost the same color, making it
difficult to reliably determine the age of the observed structures.
The extended ionized gas emission can also contribute significantly
into surface brightness profiles (Papaderos et al 2003).
Star number counts are the preferable method for studying galactic
outskirts due to the extremely low surface brightness of these
elusive components. Single-star photometry also allows us
to eliminate young stars (located mainly in the thin disk) and background
objects (e.g., Tikhonov 2002; Drozdovsky, Tikhonov \& Schulte-Ladbeck 2003;
Aparicio 2003).

Our goal here is to extend the study of stellar thick disks and halos
over a set of three spiral galaxies outside the Local group, M~81, NGC~55
and NGC~300. These galaxies provide contrasting environments, and an
opportunity to examine the stellar populations on larger galactic scales.
In particular we address the following two issues: the spatial
distribution of different stellar populations in the outskirts of spiral
galaxies and the metallicity gradient of these populations.
Brief information about the galaxies is given in Table~1.
Each of these objects has several fields imaged with
HST, situated at
various distances from the center. The high spatial
resolution of HST  allows us to perform single-star photometry even
in crowded fields. The availability of images at
various galactocentric distances allows us to
reconstruct the behavior of stellar density along
the galactocentric radius. Since thick disks and halos
consist of old stars --- red giants and, to a lesser
degree, AGB stars --- only these stars were included in our
consideration. The young stellar populations, located within the thin disks
of these galaxies, have been studied in detail (Zickgraf et al.,
1990; Georgiev et al., 1992a,b; Pritchet et al., 1987; Kiszkurno-Koziej,
1988; Pierre \& Azzopardi, 1988) and are not considered in the
current work.

A brief description of the target galaxies and justification for their
selection is given in Section~2,
followed in Section~3 by a description of the HST data
and reduction techniques.
Our results are discussed in Section~4 and summarized in Section~5.

\section{The galaxies}

\subsection{M~81}

M~81 is the gravitationally dominant member of its group, consisting of
about 30 types of galaxies (B\"{o}rngen et al. 1982, 1984;
Karachentseva et al., 1985).
The galaxy is a suitable object for this study because of
its close distance, $D=3.6$ Mpc (Freedman et al., 2001) and
small inclination, which provide an opportunity to study the geometry of
the stellar structures across the disk plane.

While the young stellar populations of M~81, populating mainly its
spiral arms, have been previously studied in numerous investigations,
little is known about the galaxy periphery.
Its optical surface brightness profile has been traced out to
limiting isophotes of $\mu_V\sim25-26^m/\sq\arcsec$ (Tenjes  et al., 1998),
corresponding to
major and minor axis diameters of about $24\arcmin\times14\arcmin$.

The radio observations of M~81 revealed hydrogen bridges from M~81
to the neighboring galaxies M~82, NGC~2976, and NGC~3077 (van der Hulst, 1978;
Appleton et al., 1981; Yun et al., 1994; Westpfahl et al., 1999;
Boyce et al., 2001).
The long gaseous filaments,
generated by galaxy interactions, are gravitationally unstable and
might be fragmenting into isolated systems, forming young Tidal Dwarf Galaxies
(TDG) (Barnes \& Hernquist, 1992; Elmegreen et al., 1993; Duc et al., 1998;
Weilbacher, 2002).
TDG candidates have been discovered in many interacting pair of
galaxies (Deeg et al., 1998; Hunsberger et al., 1996). It has been
suggested that the dwarf galaxies observed within the hydrogen bridges
of M~81 --- Ho~IX, Garland, and BK3N --- may be
TDG (Miller, 1995; Flynn et al., 1999; Boyce et al., 2001).
By comparing the results of observations with theoretical isochrones,
Sakai \& Madore (2001) concluded that the age of young stars in Garland is
less than 150 Myr. The star formation history of Ho~IX, Garland and BK3N,
estimated from the single-star photometry of the HST images
(Makarova et al., 2002), confirms that the intensive star formation in these
galaxies is on the time scale of 50-150 Myrs.
In their study, however, the possible presence of outer stars from the
neighboring M~81 and NGC~3077 has not been considered.
The underlying stellar population with ages more than 1~Gyr,
observed by Makarova et al. (2002), may be more closely related to the
outskirts of M~81 and/or NGC~3077 than to the underlying stellar population
of the studied galaxies.

\subsection{NGC~55}

The spiral SB(s)m  galaxy NGC~55 (Fig.2) is a member of the Sculptor group,
consisting of
approximately 30 galaxies (Cote et al., 1997; Jerjen et al., 2000).
Since the galaxy is seen almost edge-on (see Table~1), it is a convenient
object for studying the extent of the thick disk and halo perpendicularly to
a disc plane.

We have measured the distance to NGC~55 as 2.1~Mpc, which differs
from the previous estimations by
Graham (1982) of $D=1.45$~Mpc, based on the tip of the red giant branch (TRGB)
method, and Pritchet et al. (1987) of $D=1.34 $ Mpc, using
photometry of Carbon stars.

The galaxy's neutral hydrogen disk has an angular size of about
$45  \times 12 \arcmin$ (Puche et al., 1991).
It is plausible that part
of this gas  was
driven out by the stellar winds during galaxy evolution.
There is also a reported detection of a
large diffuse structure which extends up to about 3 kpc above the disk of
NGC~55 seen in Chandra X-ray, VLT H$\alpha$, and Spitzer FIR emission
(Oshima et al., 2002; Otte \& Dettman, 1999; Engelbracht et al., 2004).

\subsection{NGC~300}

The spiral galaxy SA(s)d (NED) NGC~300 is another member of the Sculptor
group (Fig. 3).
Graham (1982) assumed that
 NGC~300 is located at the same distance as NGC~55, i.e. $D=1.45$ Mpc.
Later, Graham (1984) estimated the
distance  $D=1.65 $~Mpc using photometry of Cepheids. The present
accepted distance is $D=2.1$ Mpc, based also on the Cepheid photometry
(Freedman et al., 1992).
Butler, Martinez-Delgado \& Brandner (2004)
determined that the distance modulus estimate based on the TRGB method
$(m-M) = 26.56$ was
in good agreement with the Cepheid distance determined by
Freedman et al. (1992).
We confirm their results independently from the $I$-band TRGB method,
estimating a distance of $(m-M) = 26.50$ based on a similar
set of HST data.

The low inclination of NGC~300 (see Table~1)  allows us to investigate
both the central part of the galaxy and its periphery. NGC~300 has a large
hydrogen disk,
 $55\arcmin\times50\arcmin$ (Rogstad et al., 1979; Puche et al., 1990),
which extends further than the visual part of the galaxy. The central part of
NGC~300 has been studied in detail,
but the information about its outer stellar populations is scarce.
The brightest stars of NGC~300 have been studied spectroscopically
for the identification of blue supergiants
and determining a metallicity gradient along the galaxy disk
(Bresolin et al., 2002). Using the RGB stars
of the disk or the halo, we are able to track the surface number density and
metallicity gradients for the fainter and elder (at least a Gyr) stellar
populations.

\section{Observations, data reduction and analysis.}
\subsection{HST WFPC2 and ACS/WFC observations}

To study the resolved stellar population of galaxies,
we obtained the available HST images of M~81 (9 WFPC2 fields),
3 WFPC2 and 6 ACS/WFC regions of NGC~300 and 3 WFPC2 fields around NGC~55.
Digital Sky Survey images of these galaxies with WFPC2 \& WFC footprints are
shown in Fig.~1, 2 and 3.
The observational data are listed in Table~2, where R is the angular
galactocentric distance of the observed field in arcminutes,
ID is the HST program number and
N$_{stars}$ is the number of detected stars.
Images were reprocessed through the standard WFPC2 and ACS STScI
pipeline, as described by Holtzman et al. (1995a). After removing cosmic
rays, we performed single-star photometry with the
packages DAOPHOT and ALLSTAR in MIDAS (Stetson, 1994). These programs use an
automatic star-finding algorithm, followed by measurements of
their magnitudes via point-spread-function (PSF) fitting that is constructed
from the isolated 'PSF-stars'. For the WFPC2 data, we applied
the aperture correction from the 1.5 pixel radius aperture
to the standard $0\farcs5$  radius  aperture size for the  WFPC2 photometric
system using the PSF-stars. The F555W, F606W and F814W instrumental
magnitudes have been transformed to standard magnitudes in the Kron-Cousins
system using the prescriptions of Holtzman et al. (1995b).
For the ACS/WFC data,
we derived an aperture correction to the $0\farcs5$ (10 pixels) standard
aperture from the data. The final ACS photometry uses the latest available
photometric zero-points in the {\sl HST\/} Vegamag system as provided by
the STScI ACS team (Instrument Science Report ACS 2004-08).
The background galaxies, unresolved blends and stars contaminated by
cosmetic CCD blemishes were eliminated from the final lists, using
their characteristic ALLSTAR parameters,
$|SHARP|>0.3$, $|CHI|>1.2$ (Stetson, 1994).

\subsection{Method and star selection.}

The primary goal of this study is to determine the basic morphological
and chemical properties of the outer stellar surroundings of the target
spiral galaxies.
Using color-magnitude diagrams and stellar luminosity functions,
we separated different stellar populations and analyzed their spatial
distribution. We have compared stellar population characteristics
of available fields differentially, avoiding some the
complications of comparisons with stellar evolution models.
Because the RGB loci for old stars are far more sensitive to metallicity
than age, we use them to construct a first-order metallicity distribution
functions, neglecting the known age-metallicity degeneracy. The spectroscopy of
individual thick disk/halo stars will be needed to break the age-metallicity
degeneracy inherent to broad-band RGB colors.

Since the purpose of our research is to define  the change of the stellar number
density with increasing galactocentric distance, we should be sure that
various selection effects do not significantly  affect the final results or at
least that their influence can be corrected. For this
the following considerations were taken into account:\\
\noindent
(i) All the data for this investigation were retrieved from the HST
archive and were obtained as a part of different programs.
While this allowed us to collect far more data than would
be possible in a single primary program, it limited our research.
The observations differ in exposure time, affecting the total number of
resolved point sources.
To minimize the influence of different depth on number
of detected stars, we established a limiting luminosity
threshold from the field with shortest exposure time. The single threshold
was used to exclude fainter stars from all fields. This limit
was established to be a magnitude brighter than the photometric limiting
depth, corresponding to the detection completeness level of $\sim40\%$ derived
from the artificial star trials. We also restricted our work to observations
in $F555W$, $F606W$, and $F814W$. Being aware of different band-width
of $F606W$ and $F555W$ we intercompare the derived $V-I$ color (from
$F606W-F814W$ and $F555W-F814W$) of RGB stars
at close galactocentric distances.\\
\noindent
(ii) By removing the cosmic ray traces, we decrease their influence on
the stellar counts. This step is important
in sparse fields of halo, where the number of residual cosmic rays is
comparable to the number of faint stars. The limiting depth selection
of stars at the level of a magnitude above the photometry limit
also allowed us to minimize the possible influence
of cosmic ray cleaning on the star counting.
As a result, the reduced number of stars
increases the statistical fluctuations, but the number
density of stars is less influenced by residual traces of cosmic rays.\\
\noindent
(iii) The completeness of star detection, especially at faint levels, is
also effected by crowding of stars, which increases towards the galaxy center.
We excluded from our analysis the central galactic areas due to difficulties
with the separation of evolved bulge stellar populations from the thick disk
and halo ones. The artificial star trials have been used
to correct the number density for
the incompleteness.\\
\noindent
(iv) Due to some inclination of the observed galaxies, it is important also
to correct the measured stellar counts for the azimuthal position of the
field. While changes in thickness of the disk might influence the results,
it should not significantly change the radial gradients. \\
\noindent
(v) Another concern here is our ability to distinguish by multi-color
photometry alone a grouping of stars in the thick
disk/halo from the plethora of stars found in the galaxy's enveloping
thin disk.
Can we separate the
AGB and RGB stars between these two disks?
For the high-inclination systems, such as NGC~55, this selection can
be made based on their extraplanar distances.
There are ways to consider thin/thick disk decomposition in
low-inclination galaxies, using their spatial distribution (2D thick/thin
disk numerical decomposition) and age-metallicity selection relying
on the color of RGB stars.
Several studies have shown that thick disk/halo stars
are uniformly older and more metal poor than thin disk ones (Gilmore \&
Wyse 1985; Carney et al. 1989).
If the thick disk stars are uniformly old, then any star whose life
expectancy is brief will necessarily belong to the thin disk, regardless
of its kinematics or metallicity.
By sorting stars on the different groups according to their
life expectancies, we can assign probable membership of stars to the
thin disk or thick-disk populations.

The measurements of the stellar density along the galaxy radius and
search for the truncation radius (edge)
of the thin disk may also clarify this question.
In one of our cases (see section 4.3), the density of RGB stars decreases
monotonously ( they do not show a two-slope profile) and does not uncover
the edge of the thin disk.

We do not consider the AGB stars as a good tracer of a thick disk.
The relative density of the AGB stars in comparison with RGB stars is lower
and their evolutionary status is less certain. However, they can be
a good indicator of the edge of the thin disk.

\section{Results}
\subsection{CM Diagrams of M~81, NGC~55 and NGC~300. }

The median number of stars detected in both the $F814W$ and
$F555W/F606W$ filters in the WFPC2 fields is about 5000,
and $\sim100000$ in the ACS fields (see Table 2).
The results of stellar photometry are presented on the
CMDs (Fig.~4 and 5). The features of these diagrams resemble those of
spiral galaxies.
All the CMDs are characterized by a noticeable red
plume, while the strength of the blue plume varies a lot.
The spatial variations are apparent from the varying strengths of the
blue and red plumes, and show systematic dependence on the galactocentric
distance of the field.
The CMDs of the circumnuclear fields show evidence
of blue and red supergiants, main-sequence stars and blue-loop stars.
The outer-field stellar populations are dominated by the ``red tangle'',
which contains the red giant branch. Above this tangle,
some of the stars are asymptotic giant branch (AGB) and red supergiant (RSG)
stars.
The dashed line in all the presented CMDs indicates the position of
the tip of the red giant branch (TRGB).

\subsection{Distances}

The errors in distances of investigated galaxies do
not significantly affect our major goal to constrain the relative stellar
density distribution. In the context of distance verification, however,
it is a valuable opportunity to compare the tip of the red giant branch (TRGB)
estimations with other methods.
The absolute magnitude of the TRGB gives an estimate of distance
with a precision and accuracy similar to that of the Cepheid
method, $\leq10$\% (Lee et al., 1993; Bellazzini et al. 2001).
The $I$-band TRGB method has been
successfully applied to a wide range of dwarf galaxies, using both
ground-based and HST photometry.

To the best of our knowledge, this is first measurement of M~81 and NGC~55
distances based on this method.
As it was shown in several studies, the alternative distance estimation
relying on the Cepheid method can significantly differ
from the TRGB and other methods.
For example, in  the case of the well studied galaxy M~33
the disagreement between two methods is $\sim0\fm3$ in distance modulus,
which is a large value for a galaxy located in the Local Group
(Lee et al. 2002; Kim et al. 2002).

  In order to estimate distances, we have used one or several fields for
each galaxy, avoiding star formation regions with bright
supergiants. We preferred fields with a large number of red giants
to decrease the statistical error of the TRGB method. These
fields are S2, S3, S4, S5, S6 for M~81 (Fig.1),  S3 for NGC~55 (Fig.2) and  S1 for
NGC~300 (Fig.3). CM diagrams of these fields are shown in Fig.~4
and 5. Typical incompleteness fractions are shown in
Fig.~6 and 7 for some of the fields. The stellar luminosity function of
each field reveals a sudden discontinuity which corresponds to the
TRGB (see Fig.8).
Using the method of Lee et al. (1993), we estimate
distances to the galaxies and mean metallicity of their red
giants.
The extinction coefficients of
all galaxies are taken from the study of Schlegel et al. (1998).

We obtain the following results for the different fields
around M~81: field  S2 $(m - M) = 27.89\pm0.10$,
field  S3 -- $(m - M) = 27.89\pm0.10$, field S4  -- $(m - M) = 27.95\pm0.10$,
field S5  -- $(m - M) = 27.95\pm0.10$ and field S6  -- $(m - M) = 27.95\pm0.10$.
The metallicity of the inner fields (S2, S3, S5, S6) varies from $-0.6$ to
$-0.7$,
while the metallicity of the outer field S4 is $[Fe/H] = -0.77$.
The mean distance modulus for M~81 is $(m-M) = 27.93\pm0.04$,
corresponding to $D=3.85\pm0.08$~Mpc, which is very close to the mean
distance of 16 other dwarf galaxies in the group, $(m-M) = 27\fm9$,
determined with TRGB method (Caldwell et al. 1998, Karachentsev et al. 2002).
This indicates that the gravitationally dominant galaxy M~81 is situated near
the center of its group and there is no asymmetry in the spatial
distribution of these galaxies.

The Cepheid distance of M~81, derived by Freedman et al. (2001),
corresponds to the distance modulus of
$(m-M) = 27.75\pm0.08$. This value differs from the TRGB distance,
derived in this paper, on average $\Delta(m-M)=0\fm18$,
exceeding the estimated errors and showing no correlation with
the metallicity gradient.

 For NGC~55 (field S3) we estimate the following values: $(m-M) =
26.64\pm0.10$, $D = 2.12\pm0.10$ Mpc and $[Fe/H] = -1.25$. Our
distance estimation differs essentially from the value of Puche et
al. (1991) $D = 1.6$~Mpc, but it is in accordance with the suggestion of
Graham (1982) that galaxies NGC~55 and NGC~300 are at the same
distance.

 For NGC~300 (field S1) we find the following results:
$(m-M) = 26.50\pm0.15$, $D = 2.00\pm0.13$ Mpc, and $[Fe/H] = -0.87$.
The distance is very similar to the result of Freedman et al.
(2001), $(m-M) = 26.53\pm0.07$ ($D = 2.02$ Mpc), based on the Cepheid
photometry.

 \subsection{ Metallicity gradient of the RGB stars along galactocentric
radius.}

The existence of metallicity gradients along the galactocentric radius
has been established in the Galaxy and other nearby
spiral galaxies, on the basis of the
spectroscopy of stellar clusters (e.g., Friel \& Janes, 1993;
Rolleston et al., 2000) and H{\sc ii} regions (e.g., Shaver et al., 1983;
M{\'a}rquez et al., 2002). However, only a few galaxies are close enough to
perform metallicity measurements from the spectrum of
evolved stellar populations due to their low luminosity
(Reitzel \& Guhathakurta, 2002).
The idea to use the $V-I$ color of the RGB stars comes from its much
greater sensitivity to metallicity than it has to age, making the
mean RGB color a good stellar metallicity indicator, although there is
some degeneracy (Lee, Freedman \& Madore 1993).
The advantage of this method is that red giants
can be situated at relatively large galactocentric distances,
where surface brightness is very low and spectral methods are
below the sensitivity limits.

Among the selected galaxies
the most apparent metallicity-age gradient is observed in
M~81, based on the comparison of the mean colors of the RGB locus
for the different fields. The mean metallicity
of red giants in the S2, S3, S5, and S6 fields is  $ [Fe/H]=-0.65\pm0.03$.
For the field S4, which has a larger galactocentric distance
(Fig.~1) the mean RGB metallicity is $[Fe/H]=-0.77$.

The metallicity gradient of the galaxy NGC~300 in the fields S1 and F4
can be seen in Fig.~8. The maximum of the stellar color distribution
(mode of the distribution) traces the RGB position.
This maximum changes from $(V-I) = 1.40$ to $(V-I) = 1.48$ along
the galactocentric radius, conforming with a
metallicity increase of the RGB stars assuming a
uniform prevailing age.
  We investigated the color of RGB stars throughout
the thin disk of NGC~300 using the ACS/WFC and WFPC2 images, and
find a color gradient of RGB stars
at the edge of the thin disk (fields S1 and F4) and
near the center of the galaxy (Fig.~8). The decreasing  of
the color index towards the center of galaxy may be a combining
effect of the age and metallicity of red giants.
Thus increase in the number of the red giants near the center of galaxy
can be explained by either lower metallicity
or younger average age.
 At the edge of the galaxy, the color index decrease can
be explained only by a decrease of the metallicity with radius,
because the age of red giants cannot decrease at
the periphery without signs of star formation processes.
The color index  and the metallicity of the red giants
along the thin disk shows no significant changes.

The measurements of the edge-on galaxy NGC~55 revealed a very small
metallicity gradient.
This is probably due to the fact that measurements were
performed along  the small axis  of the galaxy, but not along the radius.

\subsection{Surface density of AGB and RGB stars in galaxies.}

Extended structures lacking young stars are being routinely discovered
in nearby galaxies of
different morphological types (e.g Grebel 1999;
Lee et al. 1999; Drozdovsky et al. 2002, Schulte-Ladbeck et al. 2002,
Tikhonov \& Galazutdinova 2002).
Using the method of star counts many interesting results have been
obtained, but many of the parameters of these stellar populations are still not
clear, such as their true spatial distribution, age and chemical parameters.
The well studied Milky Way outskirts contain mostly
old ($\approx10$ Gyr) low-metallicity stars. The radial
scale length for the Galactic thick disk is significantly larger than
that for the old/thin disk (Ojha 2001; Larsen \& Humphreys 2003).
A sharp border
between thick disk and the halo is detected in the galaxy M~33
(Guillandre et al., 1998). However, the authors  did
not measure the density gradients of red giants in the thick disk and the
halo.
The surface distribution of various stellar population in  M~31
(Sarajedini \& van Duyne 2001; Ferguson et al. 2002) revealed
that the stellar density as well as the mean
metallicity vary along the galaxy radius.
In Tikhonov (2002) we summarized results for 16 dwarf irregular
galaxies to show that red giants form a thick
disk around each galaxy. The size of such disk is 2-3 times larger
than those of visible part of the galaxy (Tikhonov, 2002). We did not
detect stars that belong to  the galaxy beyond these
disks. At the same time, our study of a more luminous irregular
galaxy IC10 demonstrated that it has an extended disk
(Tikhonov, 1999) and  a more extended halo behind the
sharp edge of this disk (Tikhonov, 2002; Drozdovsky et al., 2002).

In this paper,
we analyze a set of three nearby spiral galaxies using the classical
approach for the identification of thick disks and
halos: the need for an additional disk component when attempting to
fit single disk models to the stellar surface number density
(Burstein 1979; Gilmore \& Reid 1983; Pohlen 2002,2004;
Larsen \& Humphreys 2003).
Having only archival data limited our ability to analyze the true extent
of the outer stellar structures.
Unable to plan the mapping of the the galaxy outskirts,
we made an assumption of
axial symmetry for the disk and halo components,
ignoring possible tidal deformation.
Since galaxies have an inclination to the line
of sight, we calculated the distance from the investigated area to the galactic
center with the following equation:
  \begin{equation}
   Radius_0 = \sqrt((D_a)^2 + (D_b/\cos A)^2),
   \end{equation}
where $D_a $ is a projection of the seen distance from the
  investigated area to the galaxy center on the large galaxy axis,
   $D_b$ is the same projection on the small axis of the galaxy,
  and $A$ is the galaxy inclination angle to the line of site.
In this way, deprojected radii for all fields of M~81 were calculated and
the stellar surface number density profile was constructed.
The stellar number density for the different population of stars was
also corrected for the incompleteness of the sample
as described in section 3.2.

\subsubsection{ M~81.}

The HST/WFPC2 archive contains a plethora of data located at different
areas around M~81, from the central spiral arms
to the galaxy outskirts (see Figure~1).
For the central fields, we avoided regions of intense
star-formation
by masking them out, since star counts become
severely contaminated by young and intermediate-age stars,
smearing the RGB ``edges'' in luminosity and color.
On the basis of artificial star trials, we calculated the incompleteness
fractions for all studied fields, and corrected the measured stellar number
density for this effect.
In Figure~6 (top), we present the results of completeness tests for the
innermost and outermost fields, S3 and S4. For M~81 we used
red giants in magnitude range as follows:
$24\fm0 < I < 25\fm0$ and $25\fm5 < V < 26\fm5$. For field S3,
we found a sample completeness of 0.80 in I band and 0.35 in V band data.
In the case of field S4,
the completeness is 0.95 and 0.90, respectively.
The AGB stars are brighter than RGB ones, making the incompleteness
correction very small. To increase the statistical significance of the
results, we estimated a mean number density of stars (AGB and RGB) for each
of the three chips of WFPC2 (excluding the small FOV PC chip).
The obtained results, presented in Figure~9,
demonstrate an apparent drop of stellar number density along the
galactocentric radius.
This drop likely corresponds to the outer edge of the thick disk.
There is also a difference in surface number density gradients between
RGB and AGB stars: the relative fraction of RGB over AGB stars
increases with galactocentric radius.
The radius of the M~81 thick disk has been estimated as
$22\arcmin$ (see Fig.~9, bottom), which corresponds to 25~kpc. The available
WFPC2 images do not reach the outer part of the M~81 stellar halo.
The lower limit for the size of the halo radius is $\sim40$~kpc.
A similar extended stellar halo has been detected in
M~31 (Ferguson et al., 2002) and NGC5128 (Harris et al., 1999).
Some of the fields around M~81's disk and halo are contaminated
by the stellar populations of its dwarf satellite galaxies BK3N, Ho~IX,
and Arp's ring. It causes an excess of surface number density above
the average level at given galactocentric radius
(Figure~9, top).

The CMDs of satellite dwarf galaxies BK3N and Ho~IX (Figure~9) are
rich in AGB stars, faint blue stars but the number density of more evolved
stellar populations (RGB stars) is low, suggesting that BK3N and Ho~IX might
represent young tidal formations (Miller, 1995; Flynn et al., 1999;
Boyce et al., 2001).
The underdensity of the RGB stars make it impossible to determine distances on the
basis of the TRGB method.
The distance values obtained by Karachentsev
et al. (2002) refer to halo and disk stars of M~81, but not to BK3N and
Ho~IX. In the field of the Arp's ring both the AGB and RGB
stars are apparent (Figure~9), providing a lower limit on its
age, since RGB stars are at least a Gyr old.
The detection of some blue stars in this field demonstrates the presence of
low star formation in the ring.

\subsubsection{ NGC~300.}

There are three WFPC2 fields in the HST archive, located nearly along the
same galactocentric line, and six ACS/WFC fields (Figure~3).
It is clear (Fig.~10) that
the thick disk and the halo have different gradients of star surface density.
As in the case of M81, the stellar surface density of the thick
disk can be well fit by an exponential law.
The thick disk of NGC~300 reveals a sharp edge, but
it is impossible to determine the halo extent due to the lack of
the HST data at larger galactocentric radius.
As shown in Fig.~10, the NGC~300 stellar halo is
at least twice the size of the thick disk.
On the true-color picture of NGC~300 (MPG/ESO 2.2-m + WFI) a bluish disk
with a sharp border on its outskirts can also be seen.
All spiral arms are embedded in this disk.
Usually the size of such disks is considered to be the visual size of the
galaxy. Field S2 is situated on the
edge of this bluish disk where all variations of stellar
population can be studied. The number of blue stars (main sequence
giants and supergiants) rapidly decreases at the edge  of the thin disk
to null (see Fig.11), as do AGB stars,
but red giants (from the thick disk and halo) spread out further.
The RGB stars soon show  a significant drop in surface number, likely
indicating the truncated edge of the thick disk, so the estimated ratio of
the thick-to-thin disk diameters is just $\sim1.2$ (Figure~11).

\subsubsection{ NGC~55.}

NGC~55 is an edge-on galaxy, allowing us to investigate its stellar density
in the extraplanar Z-direction (perpendicularly to the plane of the disk).
From the analysis of three fields, we
determined the surface number density of the red giants in the range of
extraplanar distances from 2 to 7 kpc (Figure~2).
It was found that all three fields (S1, S2, S3), despite different
galactocentric radii, show the same gradient
in surface number density of RGB stars, suggesting that all
the fields are located in the thick disk (Figure~2).
At the same time, the surface number density distribution does not
show a sharp thick disk edge at least untill $\sim6$~kpc above the disk
plane (Figure~12).
Perhaps the deviation of density distribution from the straight line
(fig.12) at 6.5 kpc corresponds to the transition from the disk to the halo.
Thick disks of edge-on galaxies NGC~891, NGC~4244, IC~5052 have about the same thickness
(Tikhonov et al. in preparation).
To determine the true extent of the thick disk and to reveal the halo,
additional observations are necessary.
Using the average thick disk length-to-height ratio, $2.5:1$,
obtained from our study of Irregular galaxies (Tikhonov, 2002),
and assigning a lower limit for the thick disk of $\sim6$~kpc
we estimate the length (planar extent) for the NGC~55 thick
disk as $\geq15$~kpc.

Similarly to M~81, the AGB stars of NGC~55 and NGC~300 show larger
number density gradients relative to the RGB stars, and are almost absent
in the galaxy outskirts.
Due to the smaller statistics, the
derived surface number density gradients of
AGB stars are very uncertain for NGC~55 and NGC~300.

\section{ Discussion.}

Relying on the star number counts in the spiral galaxies
M~81, NGC~300 and NGC~55,
we examine the spatial distribution of stars in their thick disks
and halos. The emphasis in the current work is on the extended surroundings
of the galaxies, out to typically a few kpc.
The disks and halos of M~81 and NGC~300
have a similar structure, but differ in spatial sizes.
While the thick disk in
NGC~55 is apparent, we found only weak evidence for the
presence of a stellar halo within 6~kpc of this galaxy.

Comparing the inferred parameters for the thin/thick disk and halo
with other published results, we compose
a scaled 3-D model of the stellar structures of a typical spiral galaxy
(Figure~13).
The extraplanar size (height or thickness) of the thin disk depends on
the galaxy type (Ma, 2002).
The height of the thin disk in M81, NGC 55 and NGC 300 spans a range
of 0.7-1.5~kpc.
The thick disk, as seen in NGC~55 (Fig.12), has thickness of 13~kpc,
but the thick-to-thin disks diameter ratio is only $1.2 - 1.3$.
The same  size ratio was observed in M~33 (Guillandre et al., 1998).
By comparing the stellar structure  of galaxies M~81 and NGC~300 with those
of irregular galaxies (Tikhonov, 2002), we noticed that while spiral galaxies
have a small thick-to-thin disks size ratio, irregular galaxies demonstrate
much larger ratios, from 2.5 to 5. A very large RGB disk was found in
the dwarf lenticular
galaxy NGC~404 (Tikhonov et al., 2003).
It is plausible that all spiral
galaxies have not only a thick disk but also an extended halo, while
dwarf irregular galaxies have only a thick disk.
Knowing that bright irregular galaxy IC~10 has a thick disk and an extended
halo (Drozdovsky et al. 2003), we suggest that
in order to have a stellar halo a galaxy many need a mass above a
certain limit.
Combining this work with results obtained for the spiral edge-on galaxies
NGC~891, NGC~4244 and IC~5052 (Tikhonov et al. in preparation), we
also suggest that their halo might be well represented by an oblate ellipsoid
(see Fig.~13).
This elliptical shape may also explain the halo of spiral galaxy
M~31 (Zucker et al., 2004).
Relatively high values for the thick-to-thin disk thickness ratio, if confirmed
with additional work, may imply an independent origin for the thick disk.
In order to come to a statistically reliable conclusion, it will
be necessary to study the stellar periphery of a larger set of galaxies
with various masses. The addition of results from kinematic studies
can be also used to refine the model of galactic stellar outskirts.

\section{Summary.}

The study of the outer stellar edges in galaxies remains difficult.
However, the results presented here indicate that the pursuit of
deep single-star photometry, followed by recovering the type of stars
and their spatial distribution, might be successfully used to study
galactic outskirts. The star count method allows us to trace the
extended surroundings of the galaxies out to a few scale radii
and to constrain their spatial geometry with high confidence,
at sensitivities much below all other methods.

On the basis of this method, we have obtained the following major results:

a) The extended stellar thick disks and halos have been detected
   in M~81, NGC~300 and NGC~55.

b) There are clear differences between surface density gradients
of the evolved stellar populations assigned to
the thick disk and halo of these spiral galaxies, which allowed us to
detect the edge of the thick disk.

c) Having a large dataset allowed us to estimate the distances
to the galaxies with the TRGB method with high statistical confidence.
We determine a distance of $3.85\pm0.08$~Mpc for M~81, $2.12\pm0.10$~Mpc
for NGC~55 and $2.00\pm0.13$~Mpc for NGC~300, and a mean stellar
metallicity of $-$0.65, $-$1.25, and $-$0.87.

d) There are clear differences in color distributions of the RGB stars,
indicating plausible metallicity gradients in the thick disk
and halo of the studied galaxies.

Large statistical studies of this kind are crucial to understand the
morphology of nearby galaxies. It is necessary to have a homogeneous
dataset, obtained with the same equipment and uniformly analyzed.
The HST/WFPC2/ACS Archive contains deep homogeneous datasets
with unparalleled resolution that provide a unique way to analyze the
physical characteristics of the shapes of galaxies with a
high degree of confidence.

Our study would provide essential new input for theoretical models (as currently,
our understanding of the three-dimensional structure of spirals is
limited by central high-surface brightness disks),
and may be relevant for the origin of dark matter in these galaxies, as
well as evolution of the galaxies.
We view this archival study as part of a larger research project,
involving followup ground based multi-object spectroscopy
and wide-field imaging.

 \acknowledgements{ The authors would like to  thank to the Russian Foundation
for Basic Research for financial support under the grant 03-02-16344.
Data from the NASA/IPAC Extragalactic Database have
been used.}

{}
\clearpage
\begin{table}
\footnotesize
\caption{Properties of NGC~55, NGC~300, M~81 (from NED).}
\renewcommand{\tabcolsep}{2pt}
\begin{tabular}{lccc}\\ \hline
\hline
\multicolumn{1}{c}{Galaxy}&
\multicolumn{1}{c}{NGC~55}&
\multicolumn{1}{c}{NGC~300}&
\multicolumn{1}{c}{M~81}\\
\hline\\
RA (J2000)        & $00^h14^m54^s$ & $00^h54^m54^s$ & $09^h55^m33^s$\\
DEC (J2000)       & $-39\degr11\arcmin 49\arcsec$ & $-37\degr41\arcmin 00\arcsec$ & $69\degr 03\arcmin 55\arcsec$   \\
Morphological type & SB(s)m:sp   & SA(s)d     & SA(s)ab:LINER Sy1.8\\
Helio radial velocity (km/s) &  $ 129 \pm 3$ & $144 \pm 1$  & $-34 \pm 4 $\\
Diameter (arcmin)& $32.4\times5.6$& $21.9\times15.5$& $26.9\times14.1$\\
Magnitude       &  8\fm84        &  8\fm95     & 7\fm89\\
$A_V$            & 0\fm044        &   0\fm042   &  0\fm266\\
$A_I$            & 0\fm026        &   0\fm025   &  0\fm155\\
Inclination    & $85\degr$&                 $40\degr$&$ 59\degr$\\
\hline
 \end{tabular}
\end{table}
The Galactic extinction correction is by Schlegel et al.(1998).

The inclination is taken from LEDA.

\renewcommand{\baselinestretch}{1.2}
\begin{table}
\small
\caption{Observational log of HST.}
\begin{tabular}{lccccccr}\\ \hline \hline
\multicolumn{1}{c}{Galaxy}&
\multicolumn{1}{c}{Region}&
\multicolumn{1}{c}{Date}&
\multicolumn{1}{c}{Band}&
\multicolumn{1}{c}{R}&
\multicolumn{1}{c}{Exposure}&
\multicolumn{1}{c}{ID}&
\multicolumn{1}{r}{N$_{stars}$ }\\
			    \hline\\
M~81   & S1   &  17.04.1998 & F814w&  2.96  &  1000+1200      & 7909& 18164\\
       &      &             & F606w&  2.96  &  2$\times$1000  & 7909&\\
       & S2   &  26.01.1999 & F814w&  5.18  &  3$\times$1500  & 8059& 22933\\
       &      &             & F606w&  5.18  &  5$\times$1500  & 8059&\\
       & S3   &   4.06.2001 & F814w&  3.47  &  4$\times$500   & 9073& 19143\\
       &      &             & F555w&  3.47  &  4$\times$500   & 9073&\\
       & S4   &  30.06.2001 & F814w& 12.78  &  3$\times$1400  & 9086& 1611\\
       &      &             & F606w& 12.78  &  4$\times$1300  & 9086&\\
       & S5   &   1.09.2001 & F814w&  6.09  &  2$\times$1100  & 8584& 10150\\
       &      &             & F555w&  6.09  &  2$\times$1100  & 8584&  \\
       & S6   &  28.05.2002 & F814w&  9.52  &  800            & 9634& 7810\\
       &      &             & F606w&  9.52  &  2$\times$500   & 9634& \\
       & BK3N &  29.08.2000 & F814w& 10.87  &  600            & 8061& 519\\
       &      &             & F606w& 10.87  &  600            & 8061& \\
       &Ho IX &  27.06.2001 & F814w& 11.39  &  600            & 8061& 3655\\
       &      &             & F606w& 11.39  &  600            & 8061& \\
       &Arp's &  30.07.2000 & F814w& 17.61  &  600            & 8061& 956\\
       &loop  &             & F606w& 17.61  &  600            & 8061& \\
 NGC 300 & S1 &  13.09.2001  &  F814w& 7.12 &  2$\times$500  & 8584 & 4387\\
	 &    &              &  F555w& 7.12 &  2$\times$500  & 8584 & \\
	 & S2 &  02.07.2001  &  F814w& 5.99 &  2$\times$300  & 9162 & 12004\\
	 &    &              &  F606w& 5.99 &  2$\times$300  & 9162 & \\
	 & S3 &  20.06.2001  &  F814w& 12.83 & 4$\times$500  & 9086 & 2520\\
	 &    &              &  F606w& 12.83 & 4$\times$500  & 9086 &\\
         & F1 & 17.07.2002  &  F555w& 8.05 &  1080 & 9492 &  89615\\
         &    &              &  F814w& 8.05 & 1080  & 9492 &\\
         & F2 & 19.07.2002  &  F555w& 2.07 & 1080  & 9492 & 182106\\
         &    &             &  F814w& 2.07 & 1080  & 9492 &\\
         & F3 & 28.09.2002  &  F555w& 0.91 & 1080  & 9492 & 191107\\
         &    &             &  F814w& 0.91 & 1080  & 9492 &\\
        & F4 &  21.07.2002 &  F555w& 8.97 & 1080  & 9492 &53494\\
        &    &             &  F814w& 8.97 & 1080  & 9492 &\\
        & F5 &  25.12.2002 &  F555w& 5.45 & 1080  & 9492 & 128370\\
        &    &             &  F814w& 5.45 & 1080  & 9492 &\\
        & F6 &  26.09.2002   &  F555w& 6.15 & 1080  & 9492 &68730\\
         &    &             &  F814w& 6.15 & 1080  & 9492 &\\
 NGC 55  & S1 & 13.06.2001 &  F814w&  8.00 & 4$\times$500  & 9086 & 832\\
	 &     &             &  F606w&  8.00 & 4$\times$500  & 9086 & \\
	 & S2   & 15.06.2000 &  F814w&  3.59 & 2$\times$800+900&8697& 5654\\
	 &      &            &  F555w&  3.59 & 3$\times$800  & 8697 &\\
	 & S3   & 10.06.2000 &  F814w&  8.97 & 2$\times$800+900&8697& 10532\\
	 &      &            &  F555w&  8.97 & 3$\times$800  & 8697 &\\
 \hline
 \end{tabular}
\end{table}
\renewcommand{\baselinestretch}{2.0}
\clearpage
\begin{figure}[hbt]%
\includegraphics[angle=0, bb= 37 50 572 664,clip]{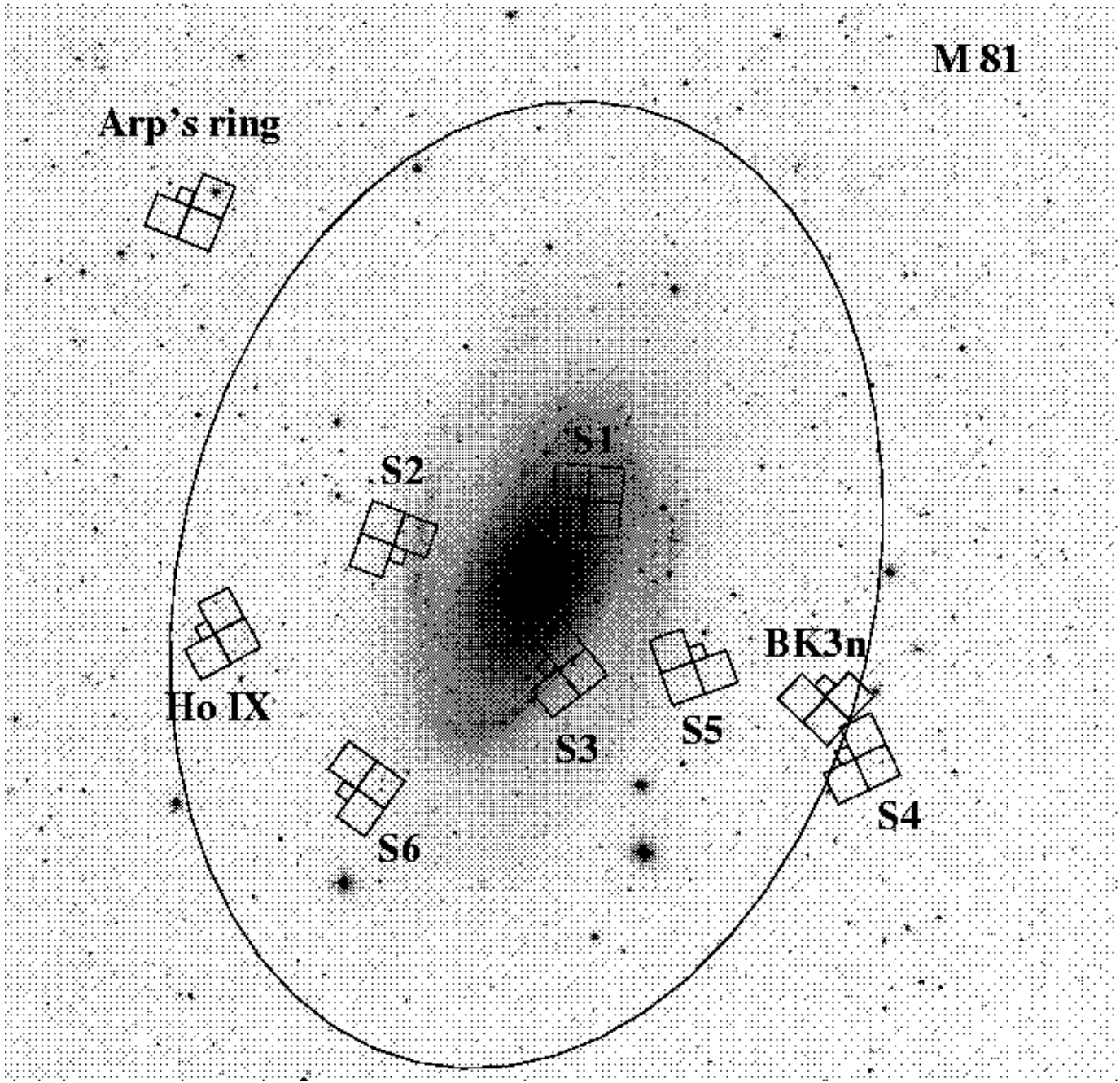}%
\caption{DSS-2 40$\arcmin\times40\arcmin$ image of M~81 with HST/WFPC2
footprints superposed, indicating the 9 regions (S1, S2, S3, S4, S5, S6, BK3N, Ho IX, Arp' ring) observed.
The edge of the thick disk of red giants is marked by an ellipse. }
\end{figure}%
\begin{figure}[hbt]%
\includegraphics[angle=0, bb= 37 50 572 664,clip]{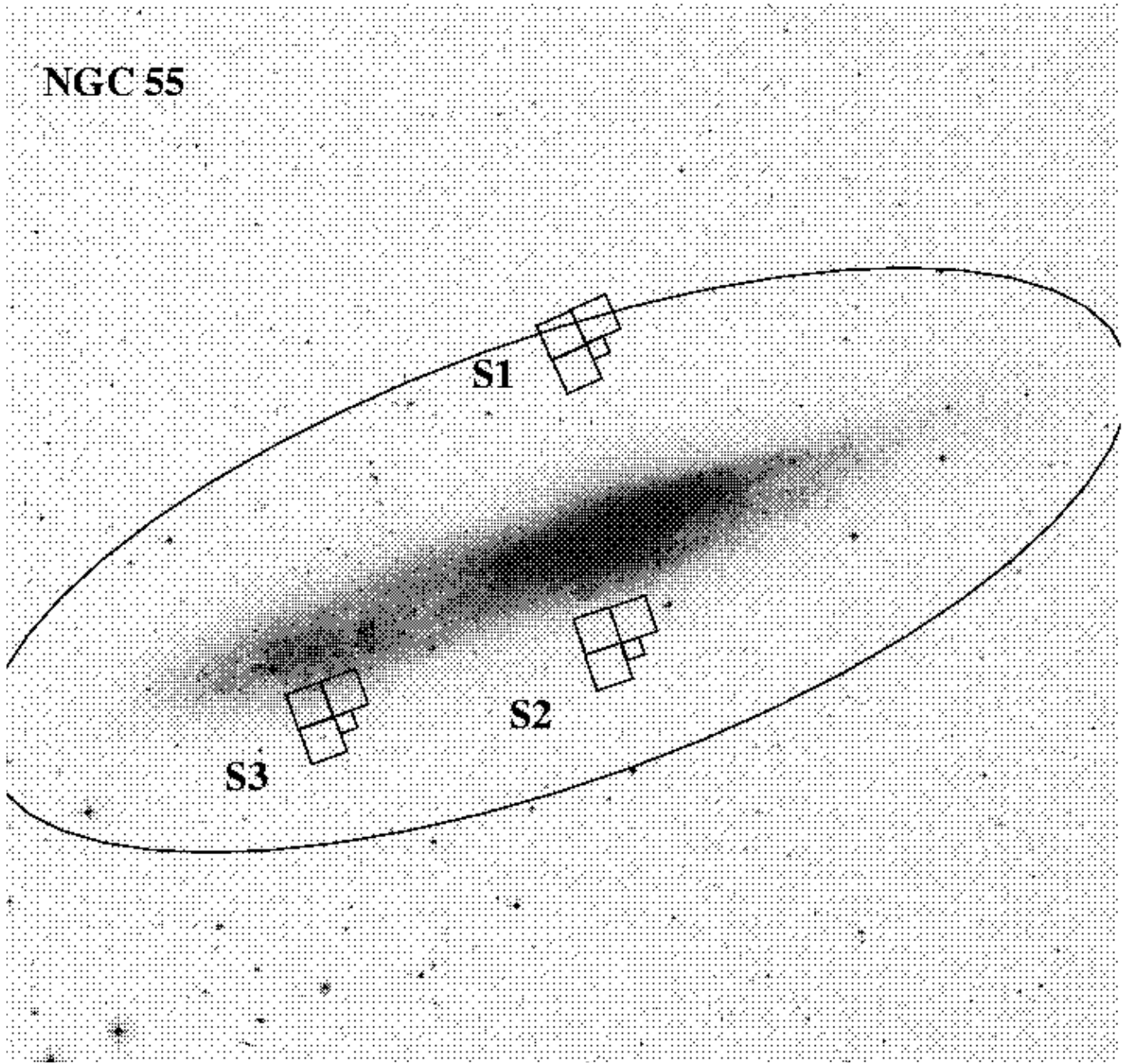}%
\caption{DSS-2 40$\arcmin\times40\arcmin$ image of NGC~55.
The location of the HST/WFPC2 fields (S1, S2, S3) is
indicated. The edge of thick disk of red giants is shown by the ellipse. }
\end{figure}%
\begin{figure}[hbt]%
\includegraphics[angle=0, bb= 17 17 570 665,clip]{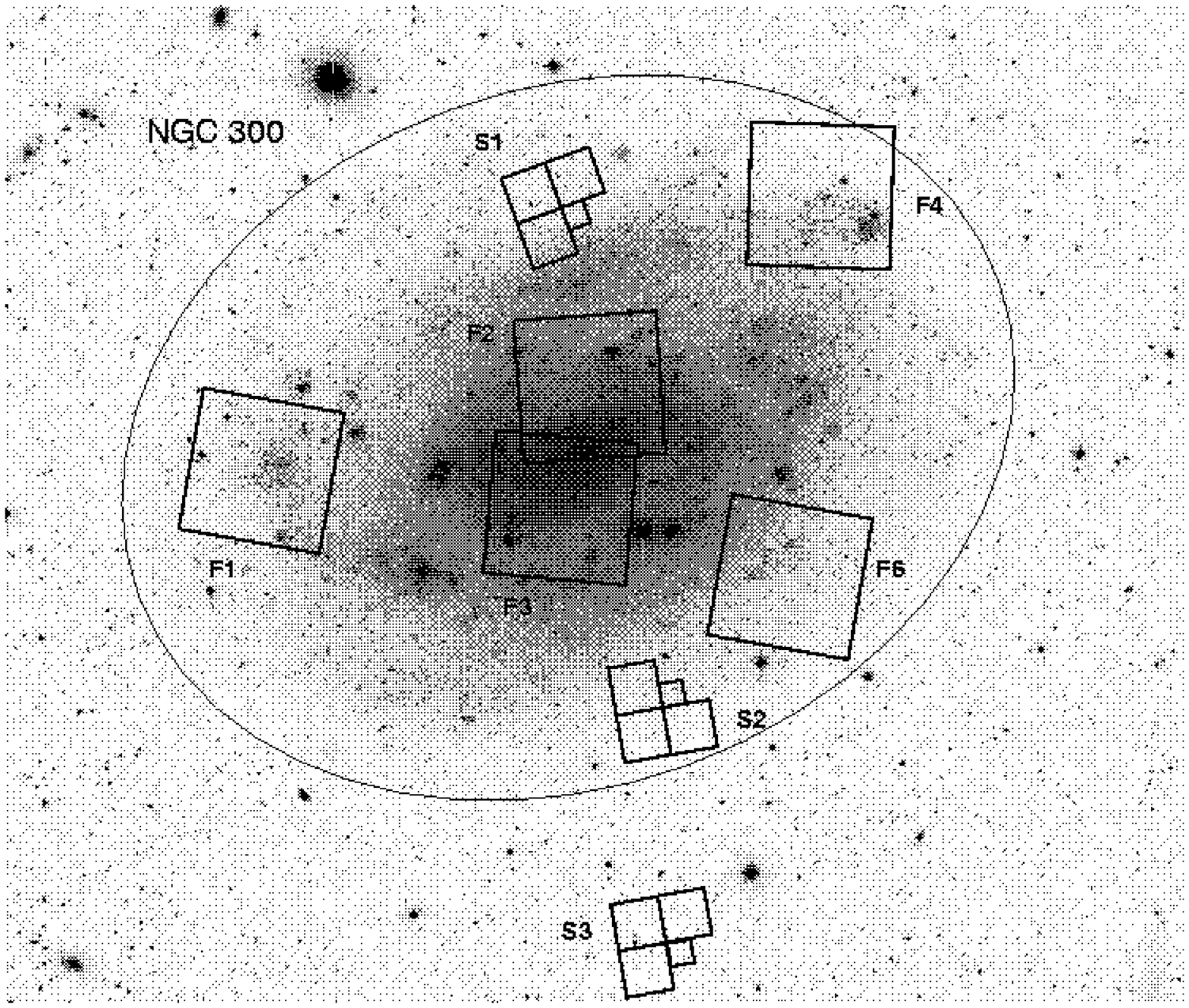}%
\caption{34$\arcmin\times33\arcmin$ WFI image of NGC 300 obtained with the
MPG/ESO 2.2m telescope.
The WFPC2 and ACS/WFI footprints are indicating 9 studied fields:
S1, S2, S3, F1, F2, F3, F4, F5, F6). The edge of the stellar thick disk
is marked by the ellipse.}
\label{f:N300_ima}
\end{figure}%
\begin{figure}[hbt]
\includegraphics[angle=270, bb= 40 30 568 547,clip]{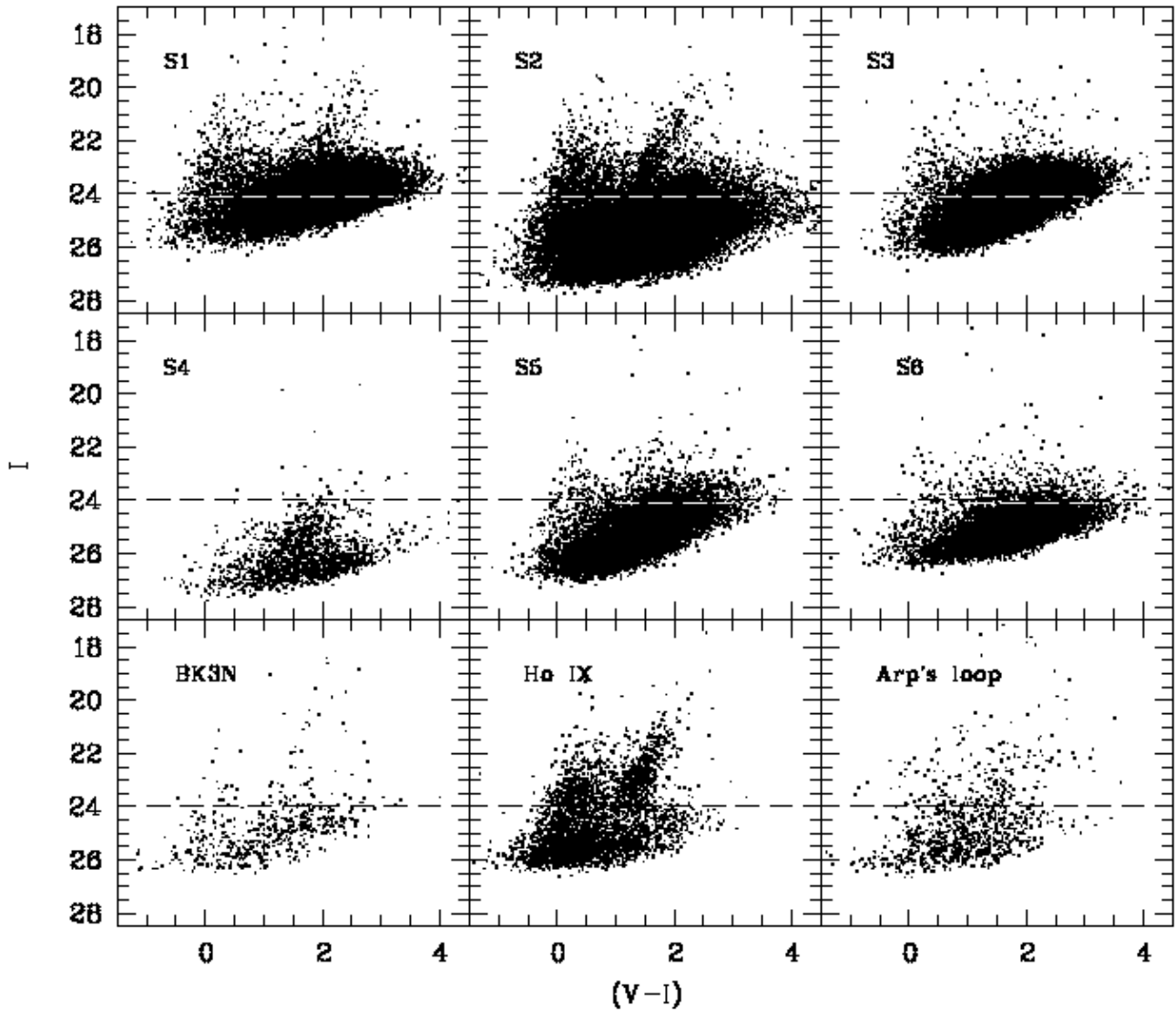}%
\caption{[$(V - I), I$] Color-Magnitude diagrams of different fields
of M~81.
The dashed line shows the position of the TRGB. Spatial variations in the
stellar content are immediately apparent from the varying strengths of the
blue and red plumes.}
\end{figure}
\begin{figure}[hbt]
\includegraphics[angle=270, bb= 40 30 560 555,clip]{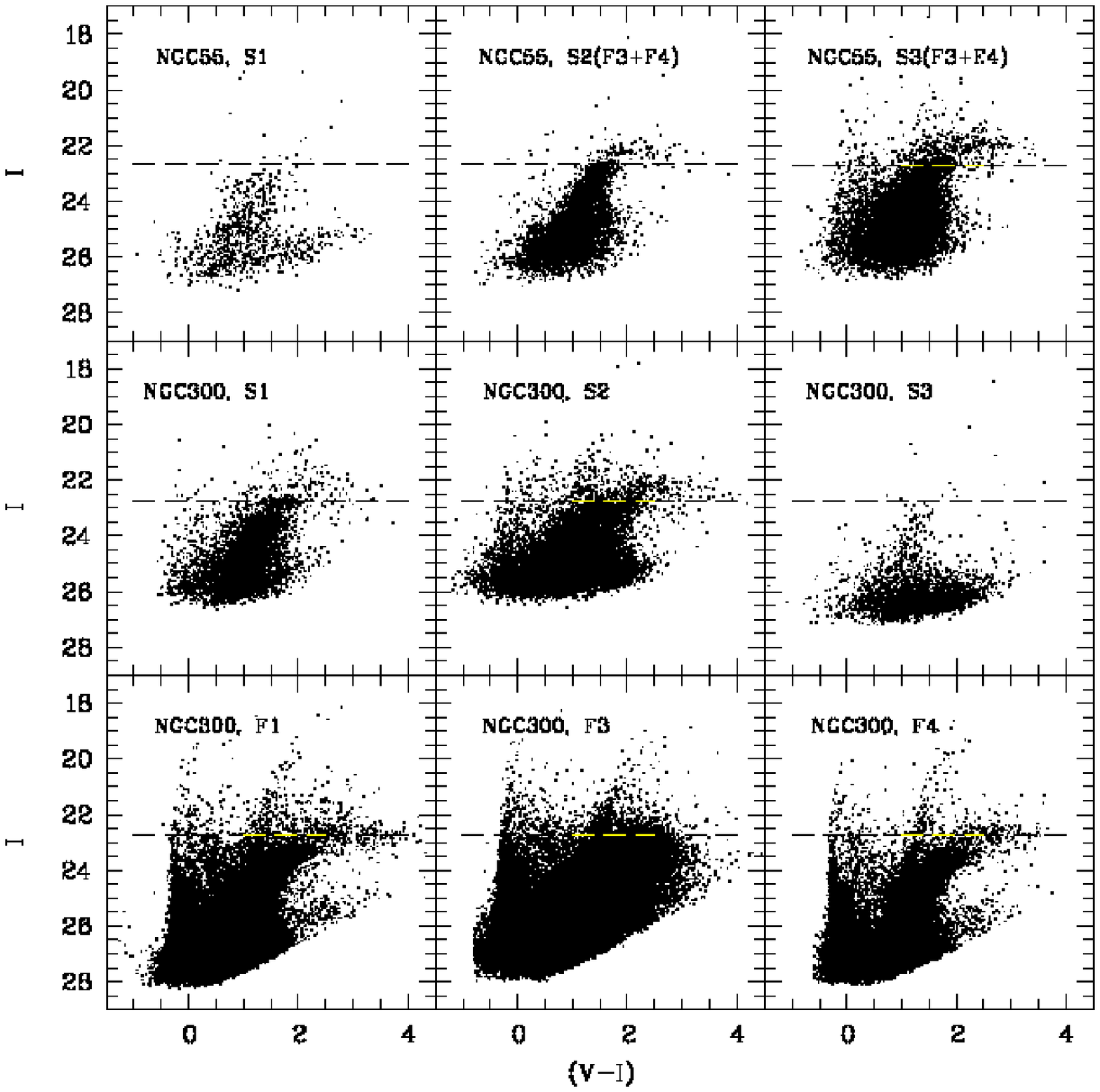}%
\caption{{\it Top. } The [$(V - I), I$] CMDs of different WFPC2 fields of
NGC~55 and NGC~300. The dashed line shows the position of the TRGB.
{\it Bottom panels} show the ACS/WFC CMDs of NGC~300 in the HST Vegamag system.}
\end{figure}
\begin{figure}[hbt]
\includegraphics[angle=0, bb= 70 320 470 800,clip]{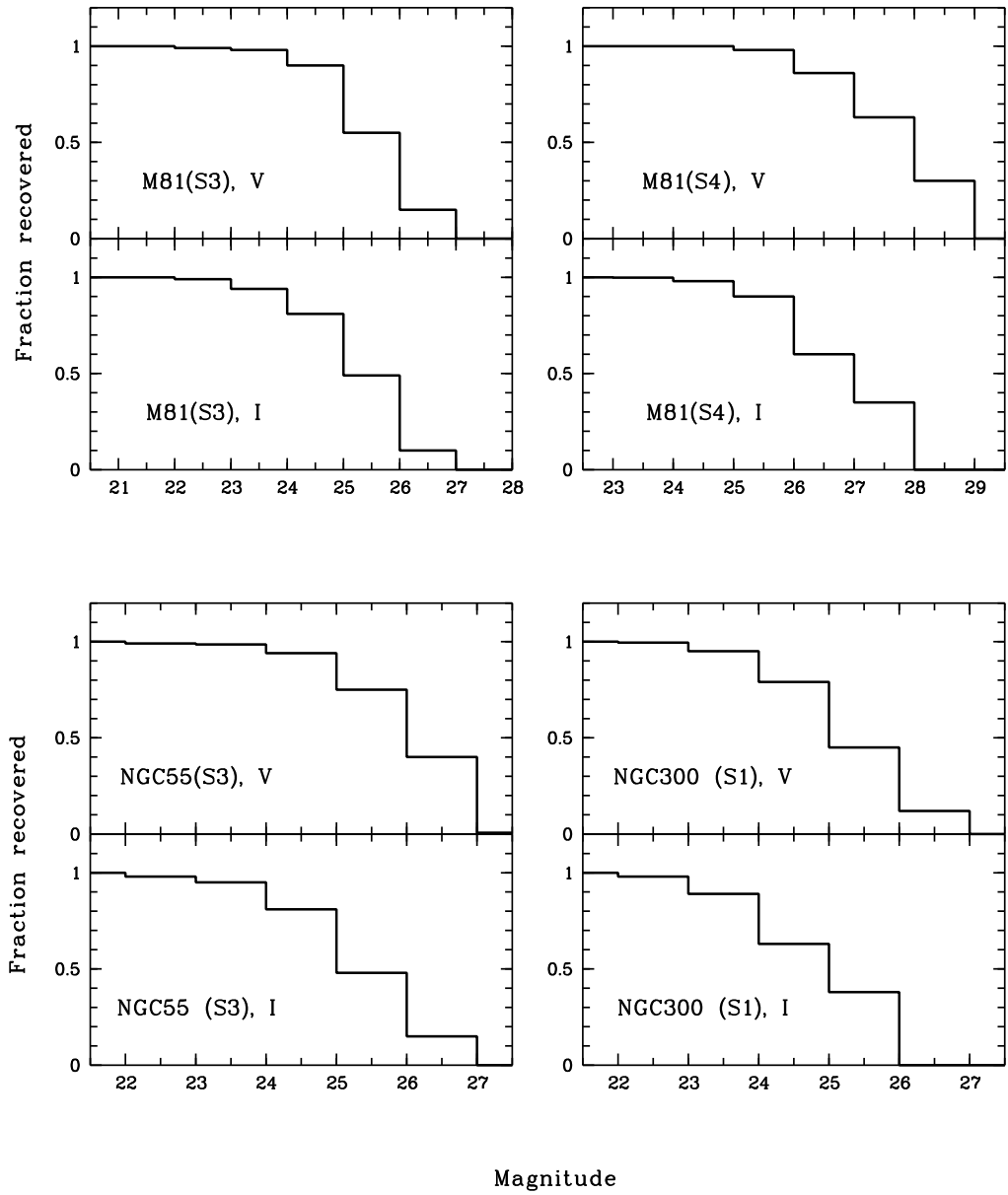}%
\caption{{\it Top.} Completeness levels of the WFPC2 photometry of two M~81
regions (S3 and S4) based on artificial star tests.
{\it Bottom panels.} Completeness levels for NGC~55 (Field S3) and
NGC~300 (Field S1).}
\end{figure}
\begin{figure}[hbt]
\includegraphics[angle=0, bb= 76 241 504 836,clip]{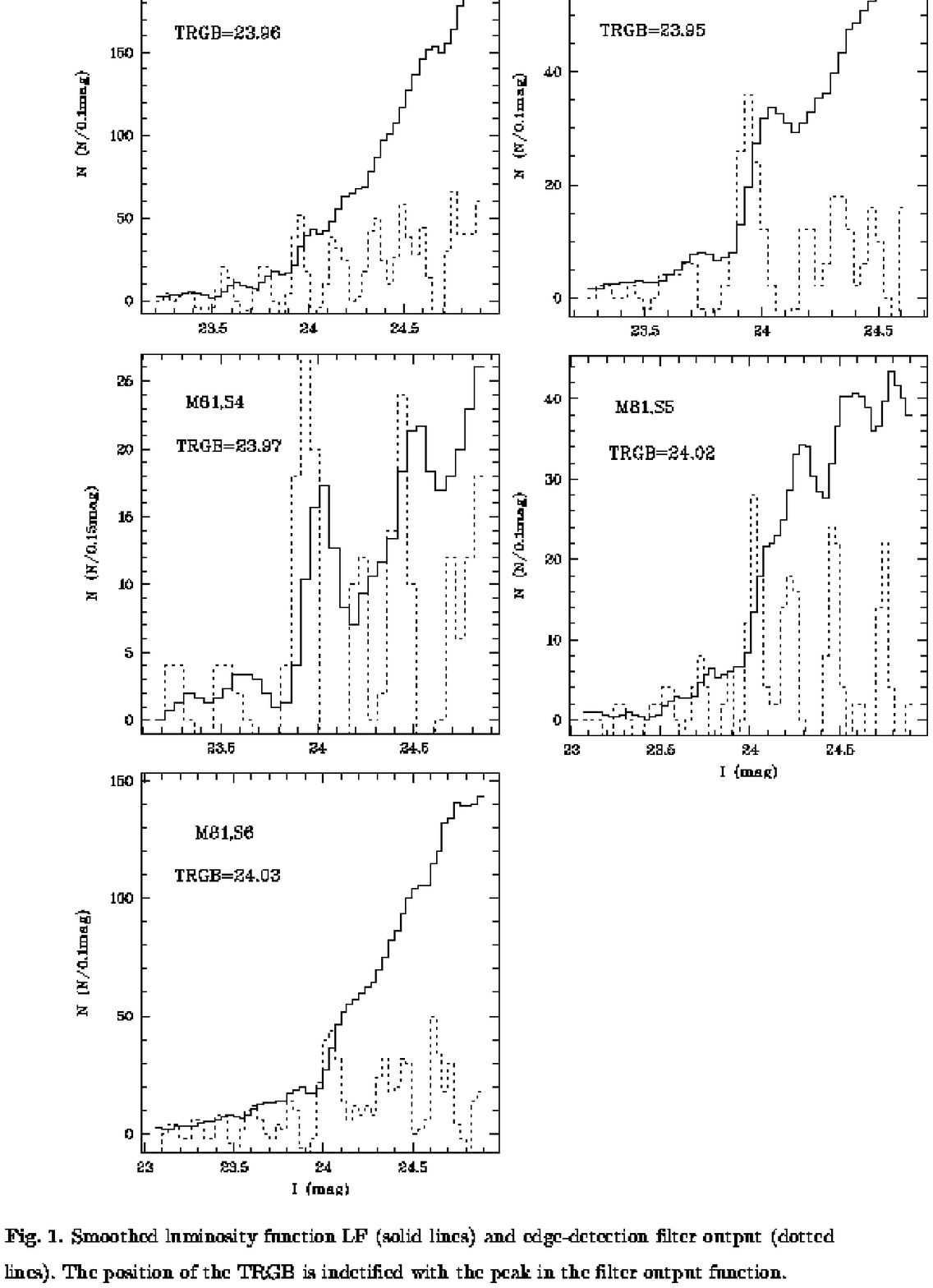}%
\caption{ Smoothed luminosity function LF (solid lines) and edge-detection
Sobel-filter output
(dotted lines) for different fields of M81. The position of the TRGB
corresponds to the peak of the Sobel-filter.}
\end{figure}
\begin{figure}[hbt]
\includegraphics[angle=0, bb= 76 380 460 815,clip]{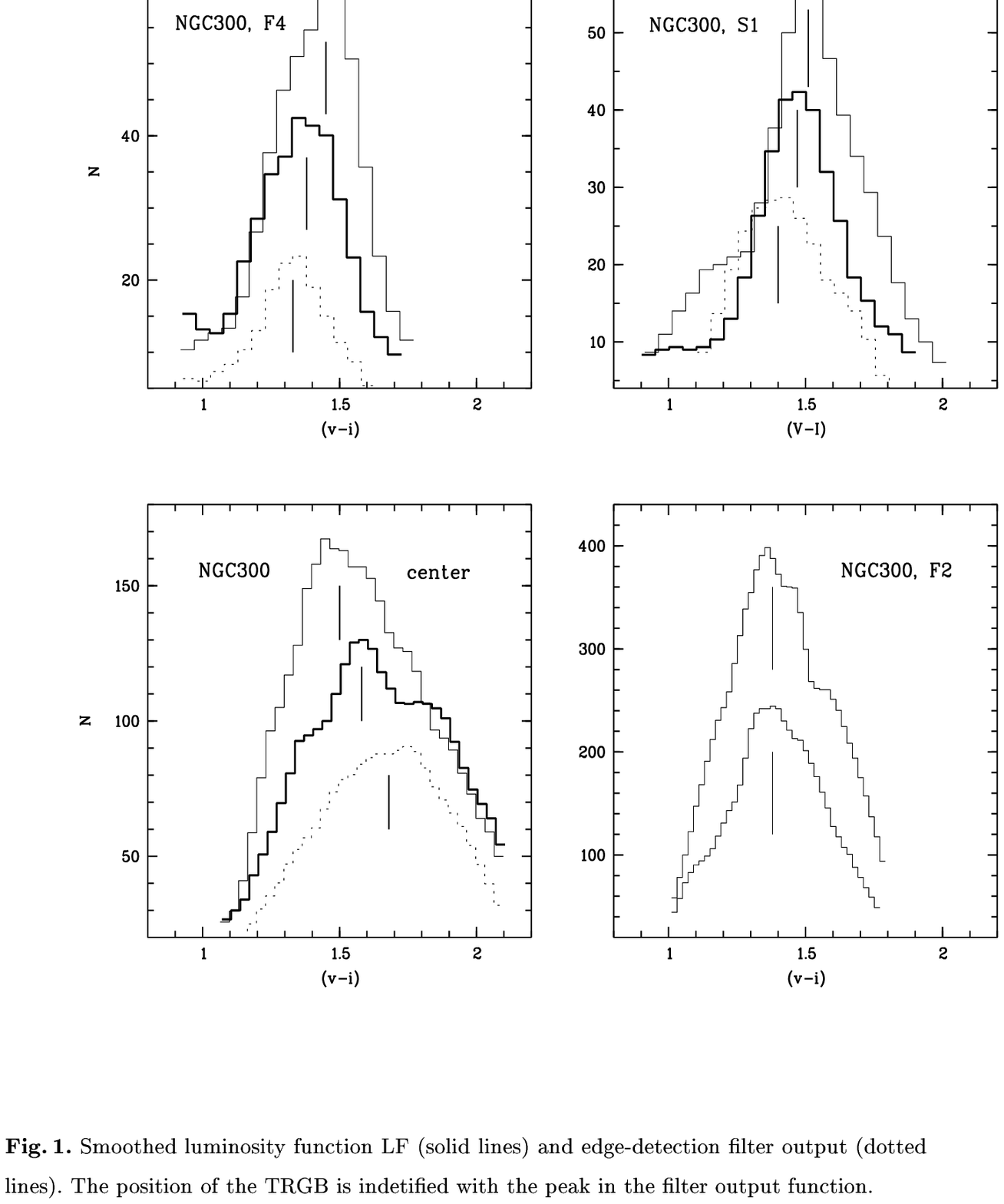}%
 \caption{The distribution of the
 $(V-I)$ (for WFPC2 data) and Vegamag $(v-i)$ (ACS/WFC) color of the RGB
 stars from 0.3 to 0.7 magnitudes below the tip of the RGB along the
 galactocentric radius for the different fields of NGC~300.
 Fields F4 and S1 are located at the edge of the galaxy thick disk,
 field F3 are near the galaxy center and field F2 is situated in the main
 galaxy body. The outer fields clearly
 demonstrate a systematic color shift along the galactocentric radius. In the
 center field F3 the decrease in color of RGB might be due to
 the effects of age, metallicity and reddenning.
 The results for the main body field F2 do not show a change of RGB
 color along the galactocentric radius.
 The dashed line corresponds to a more distant part of the chip.}
\end{figure}
\begin{figure}[hbt]
\includegraphics[angle=0, width=10cm, bb= 70 100 380 690,clip]{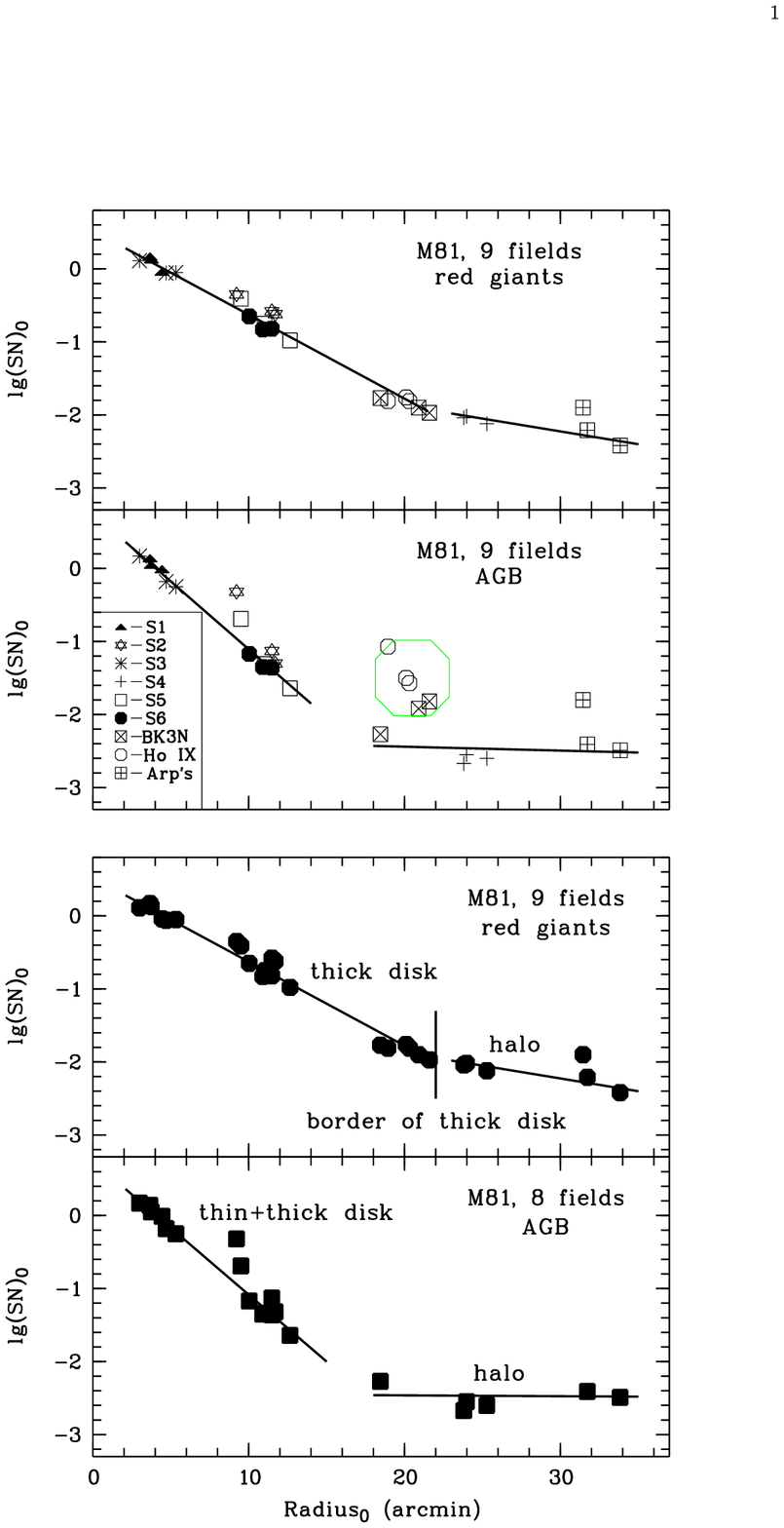}%
 \caption{{\it Top.} The surface density distribution, $(SN)_0$, for RGB
 and AGB stars, along galactocentric radius of M~81, corrected for the
 inclination.  Star counts have been corrected for the incompleteness
 based on the artificial star trials.
 The absence of data for $13\arcmin <Radius_0< 18\arcmin$
 is due to insufficient images in these regions. The deviation of some points
 from the average level of the
 stellar density is a result of the combination
 of M~81 disk/halo evolved stars with the stars of its satellites
 BK3N, Ho IX, and Arp's ring. {\it Bottom panels.} show the stellar surface density distributions of M~81
 corrected for contamination of satellite galaxies. This allow us to determine the edge of the thick
disk. The difference in gradients of AGB and RGB stars surface density is apparent.}
\end{figure}
\begin{figure}[hbt]
\vbox{\includegraphics{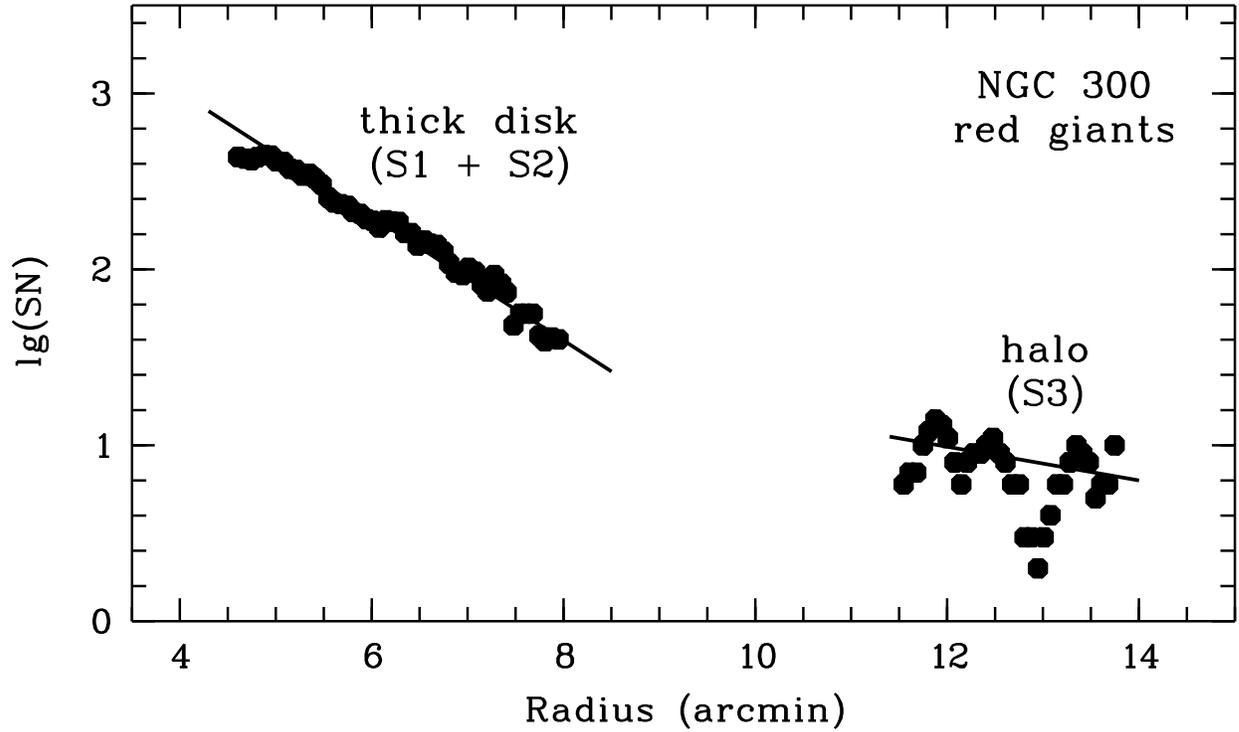}}\par
\vspace{14.0cm}
\caption{The surface density distribution (SN) of RGB stars along the
galactocentric radius. Similarly
to M~81, the stellar density distribution suggests two different gradients,
which we attribute to the thick disk and halo components.}
\end{figure}
\begin{figure}[hbt]
\vbox{\includegraphics{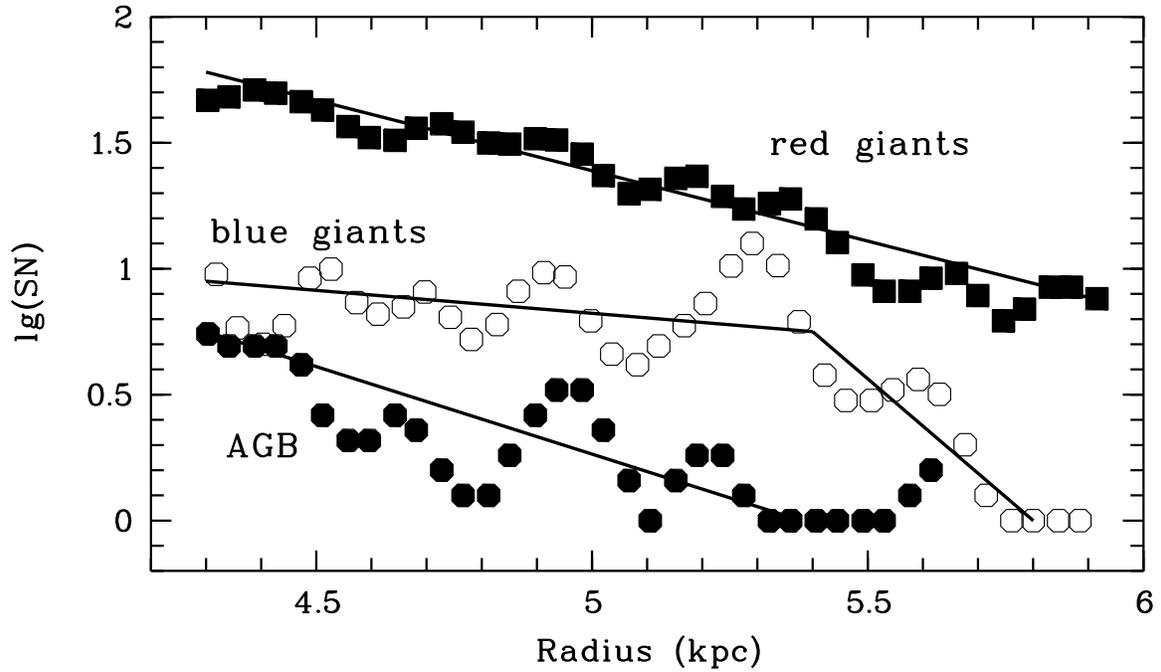}}\par
\vspace{16.0cm}
\caption{The surface density distribution of RGB, AGB and blue stars in the
edge of thin disk of NGC~300 (field S2).
The drop in the surface number density of blue stars at the galactocentric
radius of 5.7~kpc suggests that we reach the edge of thin disk.
The red giants of the thick disk extend to a larger radii. Additional
observations targeted at studying parts with larger galactocentric distances
are necessary to determine the true size of the stellar thick disk/halo at
this galaxy.}
\end{figure}
\begin{figure}[hbt]
\vbox{\includegraphics{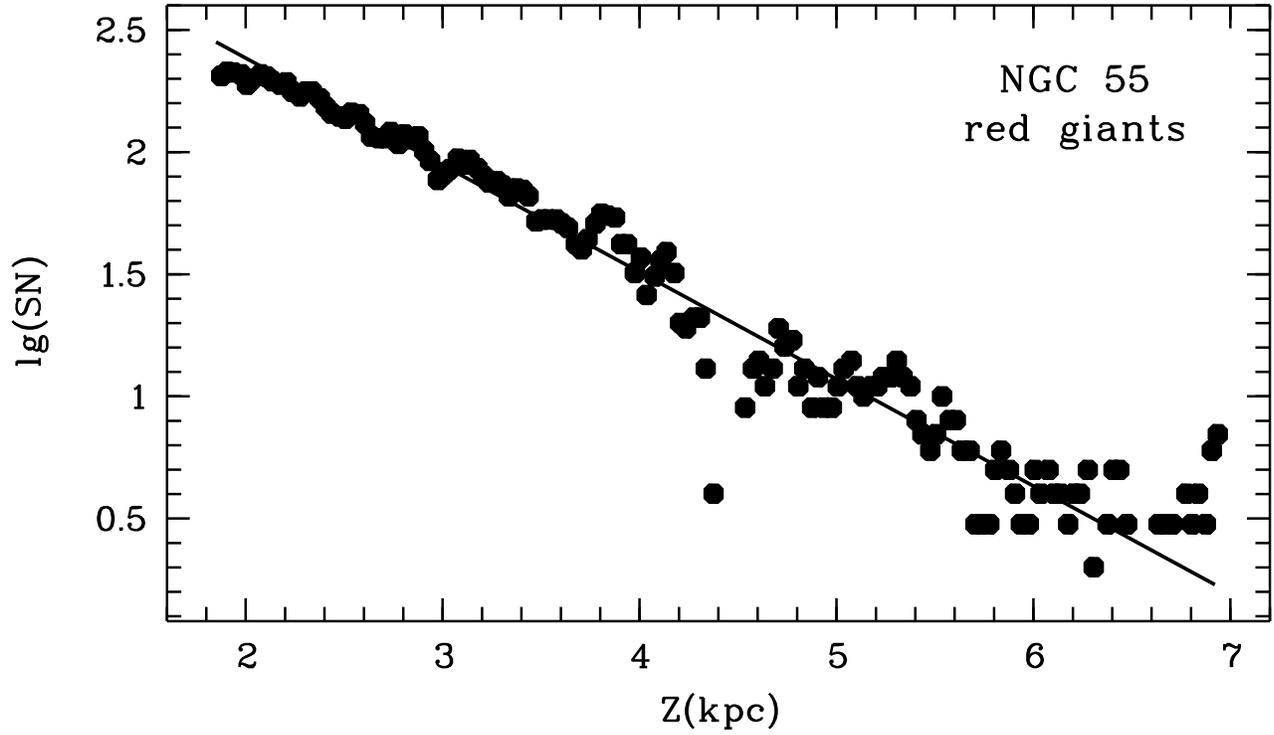}}\par
\vspace{14.0cm}
\caption{The surface density distribution of RGB stars in NGC~55
(S1, S2, S3 fields) perpendicular to the galactic plane. The thick disk
component is dominant with a possible
trace of the halo cab at $Z>6$~kpc.}
\end{figure}
\newpage
\begin{figure}[hbt]
\includegraphics[angle=0, width=110mm , bb= 0 0  400 330,clip]{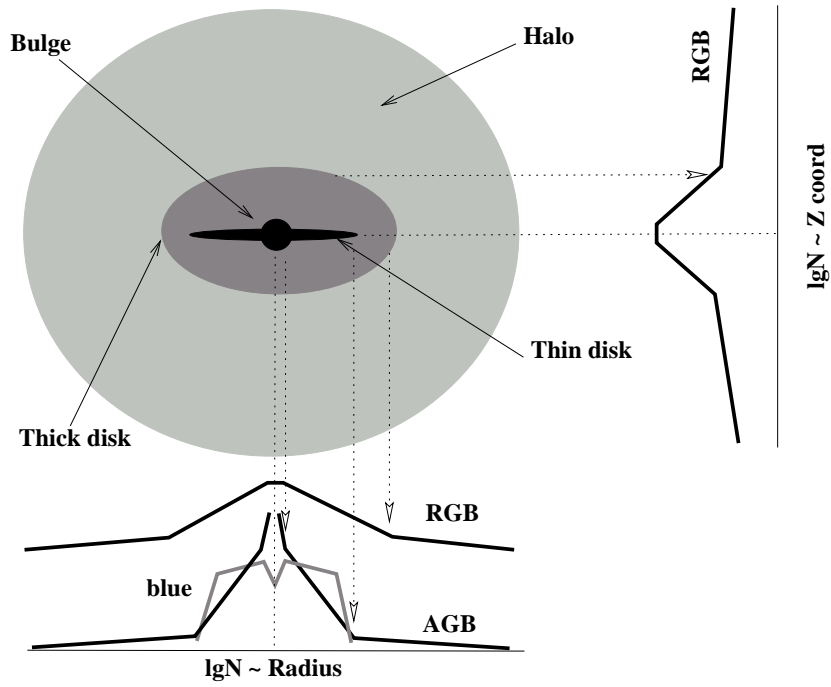}%
 \caption{A scaled 3-D representation of the stellar components of
 a typical spiral galaxy and results of their projection as a stellar number
density for a face-on (bottom plot) and edge-on (right) galaxy.
The thickness of the thin disk is from Ma(2002).
The relative sizes of thick to thin disks are from our investigations of
M81, NGC300. This results is also in agreement with M~33 by Guillandre(1998).
The relative size of stellar halo in M~81 and NGC~300 is shown as a lower
limit.}
\end{figure}
\end{document}